\documentclass[10pt,twocolumn]{article}

\usepackage[margin=0.75in]{geometry}
\usepackage{times}
\usepackage{graphicx}
\usepackage{booktabs}
\usepackage{multirow}
\usepackage{amsmath}
\usepackage{amssymb}
\usepackage{tcolorbox}
\usepackage{url}
\usepackage{hyperref}
\usepackage{natbib}

\date{}

\begin{document}

\title{\textbf{Exploring Automated Vulnerability Identification in JavaScript Code Using Large Language Models}}

\author{
\centering
\begin{tabular}{ccc}
Manit Kaushik & Ishir Bhardwaj & Pranav Gupta \\[-1pt]
{\normalsize\texttt{manit22277@iiitd.ac.in}} &
{\normalsize\texttt{ishir22223@iiitd.ac.in}} &
{\normalsize\texttt{pranav22364@iiitd.ac.in}} \\[6pt]
\multicolumn{3}{c}{
\begin{tabular}{cc}
Dr. Pankaj Jalote & Dr. Arun Balaji Buduru \\[-1pt]
{\normalsize\texttt{jalote@iiitd.ac.in}} &
{\normalsize\texttt{arunb@iiitd.ac.in}}
\end{tabular}
} \\[6pt] \\
\multicolumn{3}{c}{Indraprastha Institute of Information Technology Delhi} \\
\multicolumn{3}{c}{New Delhi, India}
\end{tabular}
}

\maketitle

\begin{abstract}
JavaScript powers approximately 98.8\% of all websites, making vulnerabilities in its code a significant security risk, yet existing detection approaches such as Static Application Security Testing (SAST) tools often fail to identify many real-world vulnerabilities when applied to isolated code snippets. This paper presents an empirical study of Large Language Model (LLM)-based vulnerability identification for JavaScript programs, evaluating three LLM families (Gemini 1.5 Flash, GPT-4o Mini, DeepSeek-R1-Distill-Llama-8B) across multiple prompting strategies (zero-shot, chain-of-thought, few-shot) and fine-tuning approaches on a dataset of 1,125 JavaScript code snippets spanning five Common Weakness Enumeration (CWE) categories: Injection (CWE-74), OS Command Injection (CWE-78), Cross-Site Scripting (CWE-79), SQL Injection (CWE-89), and Uncontrolled Resource Consumption (CWE-400). Our experiments show that LLMs substantially outperform traditional SAST tools on snippet-level vulnerability identification, with a fine-tuned Gemini 1.5 Flash model achieving 60\% detection accuracy compared to near-zero performance from rule-based analyzers. We find that fine-tuning improves accuracy from 29\% to 60\%, Chain-of-Thought prompting benefits reasoning-capable models such as GPT-4o Mini, few-shot prompting is effective for polymorphic vulnerabilities such as Cross-Site Scripting, and performance varies across vulnerability categories, reaching up to 84\% accuracy for structured vulnerabilities such as SQL Injection. These results indicate that LLMs provide a practical approach for automated vulnerability identification in JavaScript code, particularly when combined with task-aligned supervision, though they should complement rather than replace existing security analysis workflows due to limited recall and uneven performance across vulnerability types.
\end{abstract}

\section{Introduction}

\subsection{Motivation and Background}

Vulnerability detection is a core component of software security, enabling the identification of weaknesses before they are exploited in production systems. The Common Weakness Enumeration (CWE) framework provides a standardized taxonomy for classifying software vulnerabilities and supports consistent analysis across programming languages and platforms \cite{mitre2024cwe}. Traditional detection approaches rely primarily on Static Application Security Testing (SAST), which uses pattern matching and rule-based analysis to identify known vulnerability signatures in source code. While SAST tools have been widely adopted in development workflows, they face practical limitations when analyzing isolated code snippets: rule-based analyzers typically require full project context to trace data flow from sources to sinks, and their precision and coverage degrade significantly when applied to incomplete code fragments \cite{owasp2024static, kluban2022,Brito_2023, keltek2024boostingcybersecurityvulnerabilityscanning, app10249119}. Recent advances in artificial intelligence have introduced Large Language Models (LLMs) as a potential alternative for vulnerability detection, due to their ability to reason over source code and capture semantic patterns without requiring explicit control-flow or data-flow graphs \cite{sheng2025llmssoftwaresecuritysurvey, zhang2024llmsmeetcybersecuritysystematic}. However, JavaScript remains comparatively underexamined in this setting, even though it powers approximately 98.8\% of websites and is frequently associated with security-critical behavior such as user input handling, authentication, and DOM manipulation \cite{kluban2022,10.1145/3468264.3473122, zenodo2024vuln}. This work focuses exclusively on automated vulnerability identification in JavaScript using LLMs, constructing a dataset of real-world JavaScript snippets spanning multiple CWE categories and evaluating three LLM families (Gemini 1.5 Flash, GPT-4o Mini, DeepSeek-R1-Distill-Llama-8B) under different prompting and fine-tuning settings, with model performance benchmarked against widely used SAST tools.

\subsection{Research Questions}

We study the following research questions to assess the viability of LLM-based vulnerability identification in JavaScript:

\begin{itemize}
  \item \textbf{RQ1:} How do LLMs perform in identifying JavaScript vulnerabilities compared to SAST tools?
  \item \textbf{RQ2:} How do prompting strategies affect detection performance across CWE categories?
  \item \textbf{RQ3:} What is the effect of fine-tuning on vulnerability identification accuracy?
  \item \textbf{RQ4:} How does performance vary across different CWE categories?
\end{itemize}

\subsection{Paper Organization}

The remainder of this paper is organized as follows. Section 2 discusses related work on LLM-based vulnerability detection and positions our contribution within the broader research landscape. Section 3 describes our experimental design and methodology, including dataset construction, model selection, prompting strategies, and fine-tuning procedures. Section 4 presents our experimental results organized by research question. Section 5 discusses threats to validity and limitations of our approach. Section 6 concludes with a summary of findings and directions for future work.

\section{Related Work}

Recent work has examined the use of Large Language Models for software vulnerability detection, with a focus on prompt design, reasoning, and model adaptation \cite{sheng2025llmssoftwaresecuritysurvey}. Gao et al.\ \cite{gao2023fargonevulnerabilitydetection} report that GPT-4 outperforms traditional deep learning models and static analyzers on Capture The Flag (CTF)-style datasets, publicly available challenge sets where participants identify intentionally embedded vulnerabilities, although performance decreases on more complex software projects. Kostina et al.\ \cite{kostina2025largelanguagemodelstext} find that LLMs outperform classical machine learning approaches across complex binary and multiclass classification tasks in vulnerability detection.

Several studies explore structured prompting strategies to improve detection accuracy and explanation quality. Ullah et al.\ \cite{ullah2023stepbystep} apply few-shot prompting and Chain-of-Thought reasoning for C/C++ vulnerabilities, while Nong et al.\ \cite{nong2024chainofthoughtpromptinglargelanguage} introduce vulnerability semantics-guided prompting, showing improvements in precision and explanation quality for Java and C++ code.

Despite these advances, prior work has limited support for JavaScript. Sheng et al.\ \cite{sheng2025llmssoftwaresecuritysurvey} identify weak cross-language generalization and scarce JavaScript evaluation as open gaps in current research. Existing datasets and experiments primarily emphasize C, C++, Java, and Python, with fewer studies examining JavaScript's dynamic execution model and frontend-oriented vulnerability patterns.

Our work contributes to this line of research by focusing specifically on \emph{automated vulnerability identification in JavaScript}. We construct a dataset of real-world JavaScript snippets and evaluate LLMs across prompting strategies, fine-tuning settings, and vulnerability categories, benchmarking against SAST tools rather than repair-oriented workflows.

\begin{figure*}[ht]
    \centering
    \includegraphics[width=0.55\linewidth]{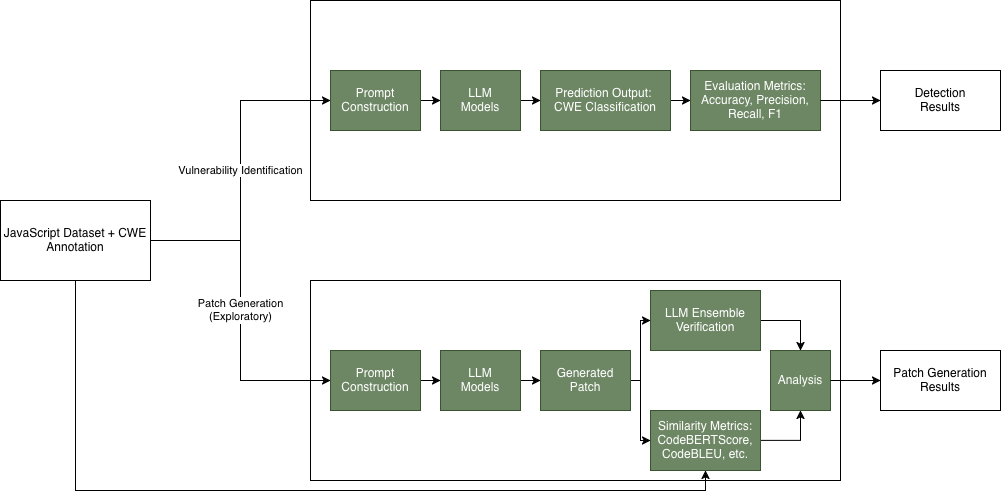}
    \caption{Experiment pipeline for vulnerability identification and patch generation.}
    \label{fig:pipeline}
\end{figure*}

\section{Experimental Design and Methodology}

\subsection{Dataset}

We curated a dataset of vulnerable JavaScript code snippets, each annotated with a Common Weakness Enumeration (CWE) identifier, from three publicly available sources: Kluban et al.'s real-world JavaScript vulnerabilities \cite{kluban2022}, CrossVul's cross-language vulnerability repository with commit-level documentation \cite{10.1145/3468264.3473122}, and Zenodo's function-level vulnerability datasets \cite{zenodo2024vuln}. We selected these three datasets because they provide real-world JavaScript vulnerabilities with verified CWE labels, commit histories that link vulnerable and patched code, and sufficient coverage across the five most frequent CWE categories in JavaScript. The number of datasets is limited to three because few publicly available JavaScript vulnerability datasets meet the requirements of verified CWE annotations, code-level granularity, and sufficient samples per vulnerability type. We selected five CWE categories based on their frequency in the source datasets and their relevance to JavaScript web applications: CWE-74 (Injection) \cite{mitre2024cwe74}, CWE-78 (OS Command Injection) \cite{mitre2024cwe78}, CWE-79 (Cross-Site Scripting) \cite{mitre2024cwe79}, CWE-89 (SQL Injection) \cite{mitre2024cwe89}, and CWE-400 (Uncontrolled Resource Consumption) \cite{mitre2024cwe400}. The final dataset contains 1{,}000 training samples and 125 test samples, with a balanced distribution of 200 training samples and 25 test samples per CWE category.

\subsection{Problem Formulation}

We treat automated vulnerability identification as a classification task over JavaScript code snippets. In the multi-class setting, the model predicts one of five CWE categories for a given snippet. In the single-class setting, the model determines whether a snippet contains a specific CWE type (binary Yes/No classification). This formulation allows us to evaluate both general vulnerability detection capability and category-specific identification performance.

\subsection{Prompting Strategies}

We evaluate three prompting strategies to assess how task framing affects LLM performance on vulnerability identification. Zero-shot prompting evaluates how well an LLM can identify vulnerabilities without examples or task-specific conditioning, serving as a baseline that reveals which vulnerability patterns are already captured in pretraining. For multi-class identification, each snippet contains exactly one of the five CWEs, and the model outputs only the CWE identifier; for single-class identification, the model responds Yes or No for a target CWE. Chain-of-Thought (CoT) prompting introduces short reasoning steps before classification \cite{wei2023chainofthoughtpromptingelicitsreasoning}, guiding the model to (1) locate potential input or attack surface, (2) assess whether the input is exploitable, and (3) map the behavior to a CWE, which allows us to study whether explicit reasoning improves identification accuracy across models, consistent with prior findings in vulnerability analysis \cite{nong2024chainofthoughtpromptinglargelanguage}. Few-shot prompting is applied only in the single-class setting, where each prompt includes one positive and one negative example retrieved using embedding similarity, allowing us to test whether in-context examples help models recognize polymorphic vulnerability patterns such as Cross-Site Scripting. We apply few-shot prompting only in the single-class setting because it requires category-specific examples, making it impractical for multi-class classification where the model must distinguish among all five CWE categories simultaneously.

\subsection{LLMs Explored}

We evaluate three LLM families to compare performance across commercial and open-source ecosystems, reasoning-oriented versus lightweight architectures, and cloud versus local execution contexts. GPT-4o Mini was selected as a commercially deployed model with strong reasoning ability at moderate cost, fine-tuned via the OpenAI Dashboard with a learning rate multiplier of 0.02 and Adam-based optimization \cite{openai2024gpt4o}. Gemini 1.5 Flash was chosen to represent a lightweight commercial architecture optimized for fast inference, fine-tuned via Google AI Studio with a learning rate multiplier of 0.002 and batch size 16 \cite{geminiteam2024gemini15unlockingmultimodal}. DeepSeek-R1-Distill-Llama-8B was included as an open-source, locally deployable alternative to support reproducibility and privacy-sensitive settings, fine-tuned using Low-Rank Adaptation (LoRA) via the Unsloth framework \cite{unsloth2024, hu2021loralowrankadaptationlarge} with rank 16 adapters on attention and MLP layers, 4-bit model loading, gradient checkpointing, mixed precision training, and AdamW 8-bit optimization \cite{deepseekai2025deepseekr1incentivizingreasoningcapability}. Models are fine-tuned on 1{,}000 labeled samples to align them with the vulnerability identification task, with standard fine-tuning using code--CWE pairs and CoT fine-tuning additionally including short reasoning traces; all models are trained for 10 epochs with batch size 16 and linear learning-rate scheduling.

\subsection{SAST Baselines}

We benchmark LLM performance against two JavaScript-oriented SAST tools chosen to reflect commonly used rule-based detection pipelines. Bearer is a security scanning platform with CWE mapping and prior use in LLM-assisted workflows, making it a suitable baseline for comparison with AI-based methods \cite{keltek2024boostingcybersecurityvulnerabilityscanning}. NodeJSScan is a JavaScript/Node.js static analyzer that supports snippet-level scanning and direct CWE outputs, enabling fair comparison with our evaluation setup \cite{Brito_2023}. Both tools operate without full-project context, aligning with our snippet-level identification task.

\section{Results}

We present experimental results organized by research question, comparing LLM-based vulnerability identification against SAST tools and evaluating the effects of prompting strategies, fine-tuning, and vulnerability category on detection performance. Performance is measured using accuracy, precision, recall, and F1-score computed over the 125-sample test set, with separate evaluations for multi-class and single-class identification tasks.

\subsection{RQ1: LLMs vs. SAST Tools}

\begin{table}[ht]
\centering
\caption{Performance Comparison: SAST Tools vs. Vanilla LLMs}
\label{tab:sast_vs_llm}
\begin{tabular}{@{}llcccc@{}}
\toprule
\textbf{Method} & \textbf{Type} & \textbf{Acc.} & \textbf{Prec.} & \textbf{Rec.} & \textbf{F1} \\ \midrule
Bearer        & SAST & 0.000 & 0.000 & 0.000 & 0.000 \\
NodeJSScan    & SAST & 0.008 & 0.200 & 0.008 & 0.020 \\ \midrule
DeepSeek-R1-8B    & LLM  & 0.080 & 0.260 & 0.080 & 0.120 \\
Gemini 1.5 Flash  & LLM  & 0.290 & 0.400 & 0.290 & 0.260 \\
ChatGPT 4o-mini   & LLM  & \textbf{0.340} & \textbf{0.430} & \textbf{0.340} & \textbf{0.310} \\ \bottomrule
\end{tabular}
\end{table}

Bearer and NodeJSScan achieve near-zero performance on snippet-level evaluation, confirming that rule-based SAST tools designed for whole-project analysis cannot effectively identify vulnerabilities in isolated code fragments \cite{Brito_2023, keltek2024boostingcybersecurityvulnerabilityscanning}. This reflects a fundamental architectural limitation: SAST tools trace vulnerability propagation through program flows, requiring project structure that snippets lack. In contrast, LLMs achieve measurable accuracy by reasoning over local semantic patterns. ChatGPT 4o-mini reaches 34\% accuracy (0.31 F1), successfully identifying roughly one-third of vulnerabilities without task-specific training, validating semantic understanding from pretraining. Gemini 1.5 Flash achieves 29\% (0.26 F1), while DeepSeek-R1-8B lags at 8\%. The substantial gap between LLMs and SAST (34\% vs. near 0\%) directly addresses RQ1, showing LLMs provide a viable alternative for vulnerability detection where traditional static analysis fails, such as code review, security audits, or educational contexts. However, modest recall indicates LLMs miss many vulnerabilities, limiting standalone use.

\begin{tcolorbox}[
  colback=gray!10,
  colframe=black!40,
  boxrule=0.4pt,
  arc=2pt,
  left=6pt,
  right=6pt,
  top=6pt,
  bottom=6pt,
  title=\textbf{Finding (RQ1)}
]
LLMs clearly outperform the SAST tools on snippet-level JavaScript vulnerability identification, but their recall is still limited, leaving many vulnerabilities undetected.
\end{tcolorbox}

These results reveal architectural trade-offs in vulnerability detection. The SAST-snippet mismatch suggests context-dependent strategies: whole-program analysis for production codebases, LLM approaches where SAST cannot operate. The 34\% baseline indicates LLMs internalized security patterns during pretraining from security discussions and code reviews in training corpora, though the accuracy-recall gap reveals high false negatives, as LLMs identify vulnerabilities correctly when detected but miss many instances.

\subsection{RQ2: Effect of Prompting Strategies}

\begin{table}[ht]
\centering
\caption{Effect of Prompting Strategies on Vanilla LLMs}
\label{tab:prompting_strategies}
\begin{tabular}{@{}llcccc@{}}
\toprule
\textbf{Model} & \textbf{Strategy} & \textbf{Acc.} & \textbf{Prec.} & \textbf{Rec.} & \textbf{F1} \\ \midrule
Gemini 1.5 Flash & Zero-Shot & \textbf{0.29} & 0.40 & \textbf{0.29} & \textbf{0.26} \\
Gemini 1.5 Flash & CoT       & 0.18          & \textbf{0.43} & 0.18          & 0.21 \\ \midrule
ChatGPT 4o-mini   & Zero-Shot & 0.34 & 0.43 & 0.34 & 0.31 \\
ChatGPT 4o-mini   & CoT       & \textbf{0.35} & \textbf{0.47} & \textbf{0.35} & \textbf{0.32} \\ \bottomrule
\end{tabular}
\end{table}

Chain-of-Thought prompting effects vary by model architecture, revealing how models process explicit reasoning (RQ2). For ChatGPT 4o-mini, lightweight reasoning guidance produces small gains: accuracy improves from 34\% to 35\% and F1 from 0.31 to 0.32, indicating structured thought processes help disambiguate similar vulnerability patterns by decomposing identification into subtasks (locating inputs, assessing exploitability, mapping to CWE). Gemini 1.5 Flash shows reduced accuracy (29\% to 18\%) despite increased precision (0.40 to 0.43), suggesting its lightweight architecture and preference for concise outputs conflict with verbose reasoning prompts, favoring precision over coverage. These results demonstrate prompting strategies must align with model capabilities: CoT benefits models trained for extended reasoning but may hinder efficiency-optimized architectures. Practitioners should validate prompting approaches per model rather than assuming universal applicability.

\begin{tcolorbox}[
  colback=gray!10,
  colframe=black!40,
  boxrule=0.4pt,
  arc=2pt,
  left=6pt,
  right=6pt,
  top=6pt,
  bottom=6pt,
  title=\textbf{Finding (RQ2)}
]
CoT prompting yields small but consistent gains for ChatGPT 4o-mini, while it reduces accuracy for Gemini 1.5 Flash, indicating that the benefit of explicit reasoning is model-dependent.
\end{tcolorbox}

\subsection{RQ3: Effect of Fine-Tuning}

\begin{table}[ht]
\centering
\caption{Vanilla vs. Fine-Tuned LLMs (Macro Accuracy and F1)}
\label{tab:finetuning}
\resizebox{\linewidth}{!}{
\begin{tabular}{@{}lcccccc@{}}
\toprule
\multirow{2}{*}{\textbf{LLM}} & \multicolumn{2}{c}{\textbf{Vanilla}} & \multicolumn{2}{c}{\textbf{Standard FT}} & \multicolumn{2}{c}{\textbf{CoT FT}} \\
\cmidrule(lr){2-3} \cmidrule(lr){4-5} \cmidrule(lr){6-7}
& \textbf{Acc.} & \textbf{F1} & \textbf{Acc.} & \textbf{F1} & \textbf{Acc.} & \textbf{F1} \\ \midrule
DeepSeek-R1-8B
  & 0.08 & 0.12
  & N/A & N/A
  & 0.27 & 0.27 \\
Gemini 1.5 Flash
  & 0.29 & 0.26
  & \textbf{0.60} & \textbf{0.60}
  & 0.22 & 0.21 \\
ChatGPT 4o-mini
  & \textbf{0.34} & \textbf{0.31}
  & 0.49 & 0.48
  & 0.30 & 0.31 \\ \bottomrule
\end{tabular}
}
\end{table}

Standard supervised fine-tuning demonstrates substantial improvements, directly answering RQ3 by showing task-aligned supervision addresses critical limitations in pretrained models. Gemini 1.5 Flash exhibits the largest gains, jumping from 29\% to 60\% accuracy (0.26 to 0.60 F1), effectively doubling identification capability. This indicates that while pretraining captures some vulnerability patterns, explicit supervision on CWE-labeled JavaScript snippets is necessary to reliably map code features to vulnerability types. ChatGPT 4o-mini increases from 34\% to 49\% (0.31 to 0.48 F1), with more modest gains likely due to stronger pretrained reasoning. DeepSeek-R1-8B, fine-tuned with CoT supervision, improves from 8\% to 27\%, demonstrating reasoning-oriented training aligns with its architecture. However, CoT fine-tuning underperforms standard fine-tuning for Gemini and ChatGPT, suggesting direct label supervision is more sample-efficient than supervising intermediate reasoning steps, practitioners should prioritize simple input-output pairs for classification tasks.

\begin{tcolorbox}[
  colback=gray!10,
  colframe=black!40,
  boxrule=0.4pt,
  arc=2pt,
  left=6pt,
  right=6pt,
  top=6pt,
  bottom=6pt,
  title=\textbf{Finding (RQ3)}
]
Standard fine-tuning on labeled JavaScript vulnerabilities yields large gains (Gemini: 29\% to 60\% accuracy). CoT-style fine-tuning helps DeepSeek but does not improve Gemini or ChatGPT beyond standard fine-tuning.
\end{tcolorbox}

Fine-tuning results reveal that pretrained models possess latent security knowledge requiring task-specific supervision to surface reliably. Gemini's 31-point improvement shows general language understanding does not automatically translate to security classification, explicit alignment through supervised examples bridges this gap. Standard fine-tuning's superiority over CoT fine-tuning suggests direct input-output mappings are more efficient for classification tasks.

\subsection{RQ4: CWE-wise Performance and Few-Shot Effects}

\begin{table*}[t]
\centering
\caption{Vanilla vs. Fine-Tuned LLMs Across CWE Categories and Prompting Methods}
\label{tab:cwe_results}
\scriptsize
\setlength{\tabcolsep}{5pt}
\renewcommand{\arraystretch}{1.0}
\begin{tabular}{@{}c l l cc cc@{}}
\toprule
\textbf{S. No.} & \textbf{LLM} & \textbf{Method} &
\multicolumn{2}{c}{\textbf{Accuracy}} &
\multicolumn{2}{c}{\textbf{F1-Score}} \\
\cmidrule(lr){4-5} \cmidrule(lr){6-7}
& & & \textbf{Vanilla} & \textbf{Fine-Tuned} &
\textbf{Vanilla} & \textbf{Fine-Tuned} \\
\midrule
\multicolumn{7}{c}{\textbf{CWE 74}} \\
\midrule
1  & DeepSeek-R1-Distill-Llama-8B & Zero Shot
   & 0.42 & 0.53
   & 0.48 & 0.59 \\
2  & DeepSeek-R1-Distill-Llama-8B & Few Shot
   & 0.53 & 0.53
   & 0.57 & 0.56 \\
3  & Gemini 1.5 Flash             & Zero Shot
   & 0.68 & 0.53
   & 0.67 & 0.57 \\
4  & Gemini 1.5 Flash             & Few Shot
   & 0.71 & 0.58
   & 0.68 & 0.62 \\
5  & ChatGPT 4o-mini              & Zero Shot
   & 0.51 & 0.66
   & 0.50 & 0.65 \\
6  & ChatGPT 4o-mini              & Few Shot
   & 0.71 & 0.78
   & 0.70 & 0.73 \\
\midrule
\multicolumn{7}{c}{\textbf{CWE 78}} \\
\midrule
7  & DeepSeek-R1-Distill-Llama-8B & Zero Shot
   & 0.51 & 0.57
   & 0.57 & 0.61 \\
8  & DeepSeek-R1-Distill-Llama-8B & Few Shot
   & 0.49 & 0.41
   & 0.56 & 0.50 \\
9  & Gemini 1.5 Flash             & Zero Shot
   & 0.79 & 0.78
   & 0.72 & 0.71 \\
10 & Gemini 1.5 Flash             & Few Shot
   & 0.79 & 0.80
   & 0.72 & 0.76 \\
11 & ChatGPT 4o-mini              & Zero Shot
   & 0.77 & 0.78
   & 0.73 & 0.74 \\
12 & ChatGPT 4o-mini              & Few Shot
   & 0.78 & 0.80
   & 0.73 & 0.73 \\
\midrule
\multicolumn{7}{c}{\textbf{CWE 79}} \\
\midrule
13 & DeepSeek-R1-Distill-Llama-8B & Zero Shot
   & 0.42 & 0.56
   & 0.49 & 0.61 \\
14 & DeepSeek-R1-Distill-Llama-8B & Few Shot
   & 0.52 & 0.40
   & 0.58 & 0.49 \\
15 & Gemini 1.5 Flash             & Zero Shot
   & 0.73 & 0.74
   & 0.72 & 0.76 \\
16 & Gemini 1.5 Flash             & Few Shot
   & 0.76 & 0.74
   & 0.70 & 0.74 \\
17 & ChatGPT 4o-mini              & Zero Shot
   & 0.46 & 0.58
   & 0.46 & 0.71 \\
18 & ChatGPT 4o-mini              & Few Shot
   & 0.76 & 0.78
   & 0.70 & 0.70 \\
\midrule
\multicolumn{7}{c}{\textbf{CWE 89}} \\
\midrule
19 & DeepSeek-R1-Distill-Llama-8B & Zero Shot
   & 0.67 & 0.59
   & 0.70 & 0.63 \\
20 & DeepSeek-R1-Distill-Llama-8B & Few Shot
   & 0.62 & 0.50
   & 0.65 & 0.56 \\
21 & Gemini 1.5 Flash             & Zero Shot
   & 0.82 & 0.78
   & 0.77 & 0.74 \\
22 & Gemini 1.5 Flash             & Few Shot
   & 0.83 & 0.82
   & 0.78 & 0.80 \\
23 & ChatGPT 4o-mini              & Zero Shot
   & 0.84 & 0.84
   & 0.79 & 0.79 \\
24 & ChatGPT 4o-mini              & Few Shot
   & 0.85 & 0.83
   & 0.81 & 0.78 \\
\midrule
\multicolumn{7}{c}{\textbf{CWE 400}} \\
\midrule
25 & DeepSeek-R1-Distill-Llama-8B & Zero Shot
   & 0.53 & 0.56
   & 0.58 & 0.60 \\
26 & DeepSeek-R1-Distill-Llama-8B & Few Shot
   & 0.48 & 0.60
   & 0.53 & 0.65 \\
27 & Gemini 1.5 Flash             & Zero Shot
   & 0.74 & 0.48
   & 0.70 & 0.53 \\
28 & Gemini 1.5 Flash             & Few Shot
   & 0.80 & 0.66
   & 0.81 & 0.73 \\
29 & ChatGPT 4o-mini              & Zero Shot
   & 0.53 & 0.66
   & 0.58 & 0.74 \\
30 & ChatGPT 4o-mini              & Few Shot
   & 0.75 & 0.79
   & 0.71 & 0.71 \\ \bottomrule
\end{tabular}
\end{table*}

Performance across vulnerability categories reveals patterns addressing RQ4 and insights into what LLMs learn about security. SQL Injection (CWE-89) emerges as easiest, with ChatGPT 4o-mini and Gemini 1.5 Flash achieving 82--84\% accuracy zero-shot, approaching practical utility. This reflects SQL injection's structured nature: vulnerable patterns involve direct string concatenation into queries (\texttt{query = "SELECT * FROM users WHERE id=" + userId}), a signature frequent in security documentation and pretraining corpora. OS Command Injection (CWE-78) shows similar performance (78--80\%), as shell command construction follows predictable patterns, indicating LLMs internalize syntactic templates for common injection vulnerabilities. Cross-Site Scripting (CWE-79) and generic Injection (CWE-74) exhibit larger variability. For CWE-79, ChatGPT 4o-mini jumps from 46\% (zero-shot) to 76--78\% (few-shot), a 30+ point improvement demonstrating value of in-context examples for polymorphic vulnerabilities. XSS patterns vary by context (innerHTML vs.\ document.write vs.\ attribute injection), and few-shot examples help recognize variations by anchoring to concrete instances. For CWE-400, few-shot improves ChatGPT and DeepSeek, though Gemini shows inconsistent behavior, suggesting lightweight architectures struggle with less-structured types.

Model-specific few-shot effects reveal architectural trade-offs: DeepSeek's few-shot performance declines in several categories, likely because its internal reasoning conflicts with externally provided examples. Gemini and ChatGPT, optimized for in-context learning, typically gain from few-shot for variable vulnerability types where examples reduce ambiguity, suggesting few-shot is most valuable for models designed for in-context adaptation.

\begin{tcolorbox}[
  colback=gray!10,
  colframe=black!40,
  boxrule=0.4pt,
  arc=2pt,
  left=6pt,
  right=6pt,
  top=6pt,
  bottom=6pt,
  title=\textbf{Finding (RQ4)}
]
LLMs detect structured vulnerabilities such as SQL Injection (CWE-89) and OS Command Injection (CWE-78) well in zero-shot, while few-shot prompting is most useful for polymorphic categories like XSS (CWE-79). The benefit of few-shot varies by model and CWE.
\end{tcolorbox}

\section{Discussion}

Our evaluation demonstrates the potential of LLM-based vulnerability identification in JavaScript while highlighting important methodological considerations. We discuss key limitations and their implications for practical deployment.

\subsection{Evaluation Context}

Our snippet-level evaluation setting favors LLMs over traditional SAST tools, which are typically optimized for whole-project analysis with interprocedural data-flow tracking. The near-zero SAST performance on isolated snippets reflects architectural differences rather than fundamental tool inadequacy, as LLMs reason over local patterns without explicit flow graphs, while rule-based analyzers require richer structural context. This suggests complementary rather than competitive roles in security workflows.

\subsection{Generalization and Dataset Scope}

Ground-truth labels are derived from publicly documented vulnerability--fix records with explicit CWE annotations. This study focuses on JavaScript and five common CWE categories; results may not generalize to other languages with different type systems or execution models, or to vulnerability types requiring complex interprocedural reasoning. We use consistent fine-tuning schedules across models to ensure fair comparison rather than optimizing hyperparameters individually, which may affect absolute performance values.

\subsection{Practical Implications}

Our results show moderate recall (34--60\%), indicating that many vulnerabilities remain undetected. This suggests LLMs are best deployed as augmentations to existing workflows, serving as rapid filters for incomplete code or exploratory contexts, rather than replacements for comprehensive static analysis. Security-critical applications should continue to employ multiple complementary detection methods.

\section{Conclusion and Future Work}

This paper investigated automated vulnerability identification in JavaScript using Large Language Models, evaluating three model families under diverse prompting and fine-tuning configurations against SAST baselines. Our study yields several contributions. First, LLMs substantially outperform SAST tools in snippet-level identification, achieving 34\% accuracy without fine-tuning versus near-zero SAST performance, establishing LLMs as viable alternatives where rule-based analyzers fail due to missing project context. Second, supervised fine-tuning produces dramatic improvements (29\% to 60\% for Gemini), revealing pretrained models contain latent security knowledge requiring task-specific alignment to surface reliably. This 31-percentage-point gain demonstrates that general language understanding does not automatically translate to security classification without explicit supervision. Third, we identify model-dependent prompting effects: Chain-of-Thought benefits reasoning-oriented models like ChatGPT but hinders efficiency-optimized architectures like Gemini, while few-shot prompting helps recognize polymorphic vulnerabilities like XSS by anchoring reasoning to concrete examples. Fourth, CWE-level analysis reveals LLMs excel at structured injection patterns (82--84\% accuracy for SQL injection) but struggle with context-dependent vulnerabilities, suggesting pretraining corpora emphasize syntactic vulnerability signatures over semantic security properties. These findings establish that LLMs offer practical value for JavaScript security in exploratory contexts, such as code review, educational settings, and rapid prototyping, where traditional SAST cannot operate effectively, though limited recall (34--60\%) necessitates integration with existing workflows rather than replacement. The substantial performance gap between vanilla and fine-tuned models indicates domain adaptation is critical for security applications, while CWE-level variance (46--84\% accuracy) suggests practitioners should validate LLM performance on vulnerability types relevant to their specific use cases rather than assuming uniform effectiveness.

Research directions emerge from our findings. Project-level evaluation would assess whether LLMs can leverage interprocedural control-flow and data-flow information beyond local pattern matching, potentially narrowing the performance gap with SAST tools in their intended whole-program analysis settings. Expanding dataset coverage to additional vulnerability classes, authorization errors, cryptographic misuse, race conditions, would reveal whether our findings generalize beyond injection-style weaknesses or whether LLMs require distinct training strategies for different vulnerability categories. Investigating hybrid approaches that combine LLM semantic reasoning with static analysis taint-tracking may improve recall while preserving precision, leveraging complementary strengths of neural and symbolic methods. Finally, longitudinal studies tracking how LLM security capabilities evolve across model generations would clarify whether vulnerability detection improves naturally through scale and diverse pretraining, or whether it requires dedicated security-oriented training data and objectives.

\bibliographystyle{plainnat}
\bibliography{sample-base}

\end{document}